\documentclass[prd,preprintnumbers,amssymb,amsmath,12pt,nofootinbib]{revtex4}

\usepackage{graphicx,multirow,subfigure}
\usepackage{bm}
\usepackage{amsmath}
\usepackage{amssymb}
\usepackage{amscd}
\usepackage{latexsym}
\usepackage{slashed}
\usepackage{color}
\usepackage{graphicx}
\usepackage{ulem}
\usepackage{color}

\begin{document}

\title{
{Exclusion and discovery via Drell-Yan in the 4DCHM}}

\date{\today}

\author{Elena Accomando}%
 \email{E.Accomando@soton.ac.uk}
\affiliation{School of Physics and Astronomy, University of Southampton, Highfield, Southampton SO17 1BJ, UK}%
\author{Daniele Barducci}%
 \email{Daniele.Barducci@lapth.cnrs.fr}
\affiliation{LAPTh, Universit\'{e} Savoie Mont Blanc, BP 110 74941 Annecy-Le-Vieux Cedex, FR}%
\author{Stefania De Curtis}%
 \email{Decurtis@fi.infn.it}
\affiliation{INFN, Sezione di Firenze, and Dept. of Physics and Astronomy, University of Florence, IT}%
\author{Juri Fiaschi}%
 \email{Juri.Fiaschi@soton.ac.uk}
 \affiliation{School of Physics and Astronomy, University of Southampton, Highfield, Southampton SO17 1BJ, UK}%
\author{Stefano Moretti}%
 \email{S.Moretti@soton.ac.uk}
\affiliation{School of Physics and Astronomy, University of Southampton, Highfield, Southampton SO17 1BJ, UK}%
\author{Claire Shepherd-Themistocleous}%
 \email{Claire.Shepherd@stfc.ac.uk}
\affiliation{Particle Physics Department,
STFC, Rutherford Appleton Laboratory, Harwell Science and Innovation Campus,
Didcot, Oxfordshire, OX11 0QX, UK}%

\begin{abstract}
 
\noindent
       
Searches for $Z^\prime$ bosons are most sensitive in the dilepton channels at hadron colliders.  Whilst finite
width and interference effects do affect the modifications the presence of BSM physics makes to Standard
Model (SM) contributions, generic searches are often designed to minimize these.  The experimental approach
adopted works well in the case of popular models that predict a single and narrow $Z'$ boson allowing these
effects to effectively be neglected. Conversely, finite width and interference effects may have to be
taken into account in experimental analyses when such $Z'$ states are wide or where several states are
predicted.  We explore the consequences of these effects in the 4-Dimensional Composite Higgs Model
(4DCHM) which includes multiple new $Z'$ bosons and where the decays of these resonances to non-SM fermions 
 can result in large widths.

\end{abstract}

\begin{flushright}
  LAPTH-Conf-038/15 \\
\end{flushright}

\maketitle

\section{Introduction}
Extra neutral massive gauge bosons, or $Z^\prime$s, are a common feature of Beyond the Standard Model (BSM) 
scenarios which can arise from general extensions of the Standard Model (SM) gauge group motivated by Grand Unified Theories 
(GUTs), Kaluza Klein (KK) excitations of SM gauge fields in models of extra dimensions,
models of compositeness and some variants of Supersymmetric models to name but a few (see Ref. \cite{Accomando:2010fz, Langacker:2008yv} and references therein). 
Typically, such objects are searched for via Drell-Yan (DY) production into two leptons: 
$pp(\bar p) \to \gamma, Z, Z^\prime \to \ell^+\ell^-$, where $\ell=e,\mu$, representing an ideal signature 
for these objects, owing to its substantial even rate, cleanliness and achievable precision. 

The latest LHC limits of relevance here placed on a variety of
$Z^\prime$ models are those obtained at around $2.5$ TeV by CMS
\cite{Khachatryan:2014fba}, with $\sqrt{\hat{s}}=7, 8$ TeV data and
full luminosity. Such limits are extracted by searching for a
resonance (so-called `bump search') in the invariant mass distribution
of di-lepton events.  These searches made the assumption that any
resonance is narrow where the width of a Breit-Wigner (BW) used to
model the signal is much smaller than the detector resolution. The results
obtained may be interpreted in a variety of models where the resonance
widths are consistent with this assumption. It was observed in
\cite{Accomando:2013sfa, Accomando:2013ita} that, by defining a cross section to be within
a particular mass window ($|M_{ll}-M_{Z^\prime}|\le 0.05\ \ E_{\rm LHC}$
where $E_{\rm LHC}$ is the collider energy) around any such resonance,
the cross sections for a wide variety of models correspond to those
predicted by the NWA. By taking this approach the difference between a
full cross section calculation including model dependent finite width
(FW) and interference effects is kept to within $O(10\% )$.
This procedure allows FW and interference effects to be treated in a
consistent way and retains the advantages intrinsic to the NWA
approach, through which model independent limits on the cross section
are derived and in turn can be interpreted as constraints on the mass
of the actual $Z^\prime$ pertaining to a specific model (i.e., the
model dependence is only contained in the di-lepton BRs of the assumed
$Z^\prime$).

All the above studies were performed in the case of single $Z^\prime$
models. The purpose of this paper is to analyze the above
phenomenology in the case of scenarios with multiple $Z^\prime$s. In
this case further challenges appear, as, in several well-motivated
theoretical models, such $Z^\prime$ states can be quite close in mass
and mix with each other. In this case the two such resonances may be close and
wide enough to appear as a broad single resonance and they may
interfere strongly with each other. The consequence of this is that
standard approach to searching for a signal resonance does not model
the deviation from SM expectations very well.

The content of our work is as follows. In Section II we introduce the multi-$Z^\prime$ model used as benchmark
and present its phenomenology, a 4-Dimensional Composite Higgs Model (4DCHM), wherein three $Z^\prime$s are active in 
the DY process. In Section III we consider the case of a single $Z^\prime$ contribution and we compare the exclusion/discovery limits computed under
NWA with the same obtained including FW and interference effects. In Section IV we discuss the case of the complete model, that is 
when multiple $Z^\prime$s are active and interfering (between themselves and with the SM). Finally in Section V we summarize and conclude.

\section{The 4DCHM}
The recently presented 4DCHM of \cite{DeCurtis:2011yx} is a scenario where the Higgs Boson arises as a 
pseudo Nambu-Goldstone Boson (pNGB) from the spontaneous breaking of a symmetry $G$ to a subgroup $H$ with the addition of the mechanism of partial 
compositeness. A minimal choice in the fermionic sector that can give rise to a finite Higgs potential, computable with the Coleman-Weinberg
technique, is assumed. 
The main characteristic of this model is the presence of a large number of heavy spin-1 resonances, due to the extra gauge symmetry $SO(5)\otimes U(1)_X$, 
and new spin-$1/2$ ones.

In the 4DCHM we in fact have the following extra gauge and matter content (alongside SM states and the Higgs boson):
\begin{itemize}
 \item 5 spin-1 neutral resonances, that are $Z^\prime$;
 \item 3 spin-1 charged resonances, that are $W^\prime$;
 \item 10 spin-$1/2$ with charge $2/3$ resonances, that are $t^\prime$;
 \item 10 spin-$1/2$ with charge $-1/3$ resonances, that are $b^\prime$;
 \item 2 spin-$1/2$ with charge $5/3$ resonances, that are $T^\prime$;
 \item 2 spin-$1/2$ with charge $-4/3$ resonances, that are $B^\prime$.
\end{itemize}

In particular, the masses of the gauge resonances are of order of $f
g_*$ for the three lightest neutral and the two lightest charged and
of $\sqrt{2}f g_*$ for the two heaviest neutral and for the heaviest
charged ones, where $f$ is the strong sector (compositeness) scale of the model
($\simeq$ 1 TeV) and $g_*$ is the common gauge coupling of $SO(5)$ and
$U(1)_X$. The widths of these extra resonances are strongly dependent
on the masses of the extra fermions: we can have a regime where the
masses of the new fermions are too heavy to allow for the decay of a
$Z^\prime$ and/or $W^\prime$ in a pair of heavy fermions, such that
the widths of the heavy gauge bosons are small, typically well below
100 GeV, and we can have the opposite configuration where the widths
of the heavy gauge bosons can become comparable with the masses
themselves (when the extra fermions are light enough).

For the purpose of a DY analysis  we have only three active $Z^\prime$ because $Z_1$ and $Z_4$ (that are the first and the 
fourth neutral gauge resonance in mass ordering) do not couple to the first two generations of quarks and leptons.

\section{Limitations of the NWA}
We now consider the direct production via DY of a $Z^\prime$ boson in the 4DCHM 
framework. Of the three active $Z^\prime$ states, to start with, we neglect the $Z_2$ and $Z_5$ states,
this in order to establish a baseline for comparison with single $Z^\prime$ scenarios.
Under this assumption we compare the results computed under NWA to those obtained following the inclusion of FW and interference effects.

\begin{figure}
\centering
\includegraphics[width=.75\textwidth]{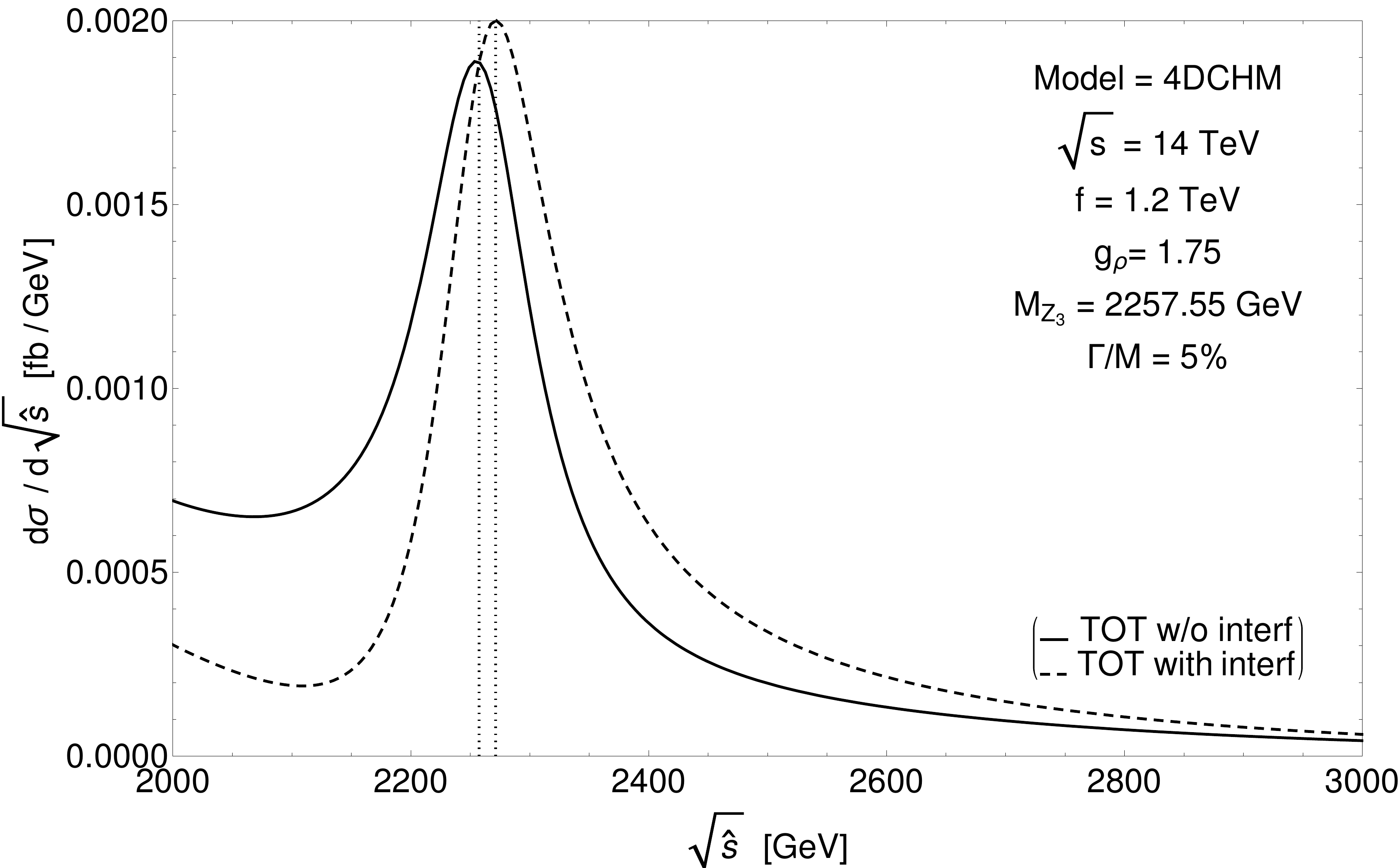}
\caption{Differential cross section distribution for a particular configuration of the parameter in the 4DCHM where only the $Z_3$ boson is active.
In the solid line we include the FW effects, in the dashed line we include both FW and interference effects. The two vertical dashed lines
represent the position of the peak in the two cases.}
\label{fig1}
\end{figure}

The first thing to note is how interference effects modify the position of the resonance peak. In fig. \ref{fig1} we show this effect for one 
point in the parameter space of the model. In this case the peak is shifted from the resonant mass pole by about 14 GeV, but there are regions in 
the parameter space of the model where the shift can grow up to 40-50 GeV.

\begin{figure}[t]
\centering
\subfigure[]{
\includegraphics[width=.45\textwidth]{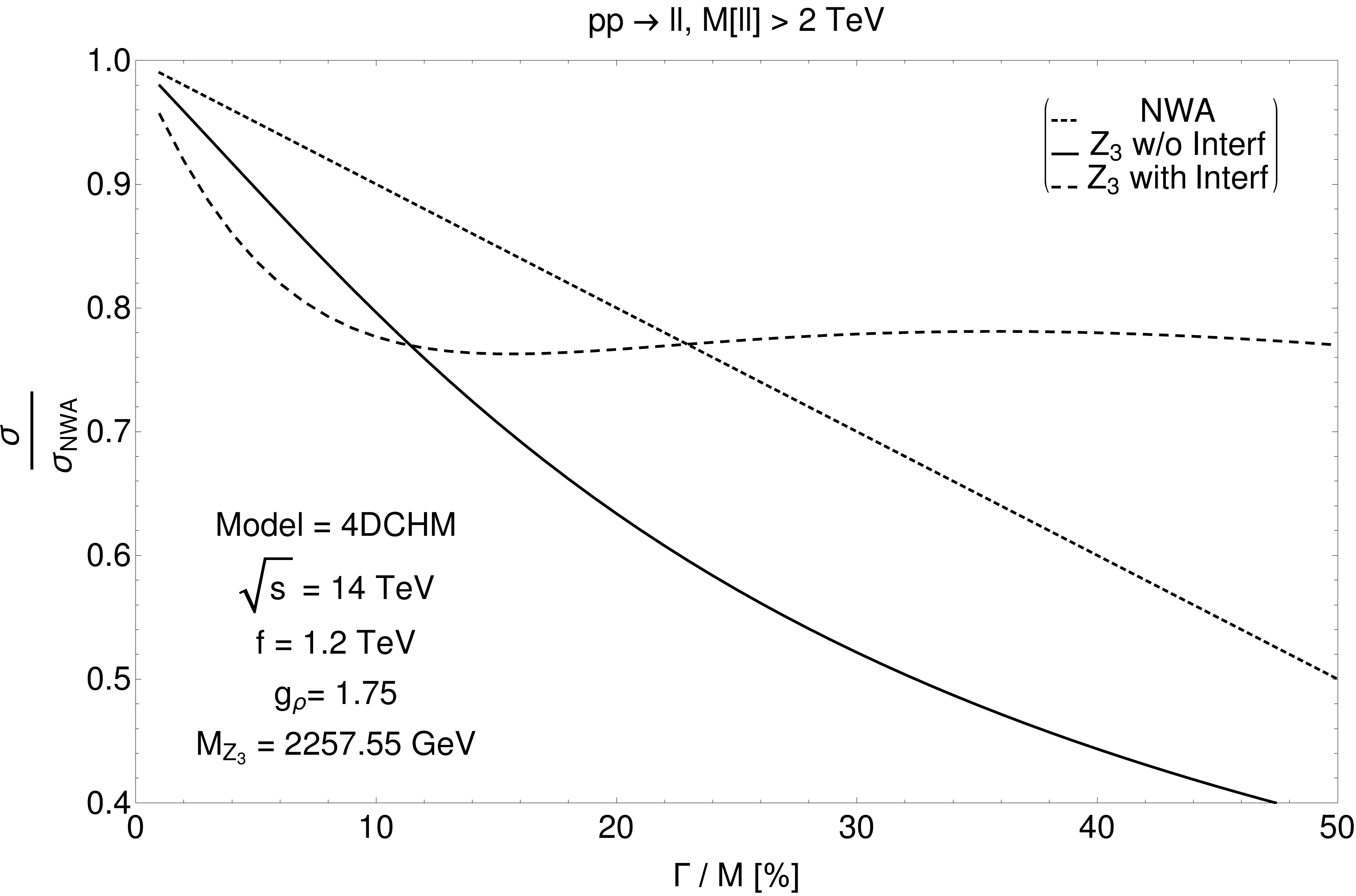}
\label{fig2a}
}
\subfigure[]{
\includegraphics[width=.45\textwidth]{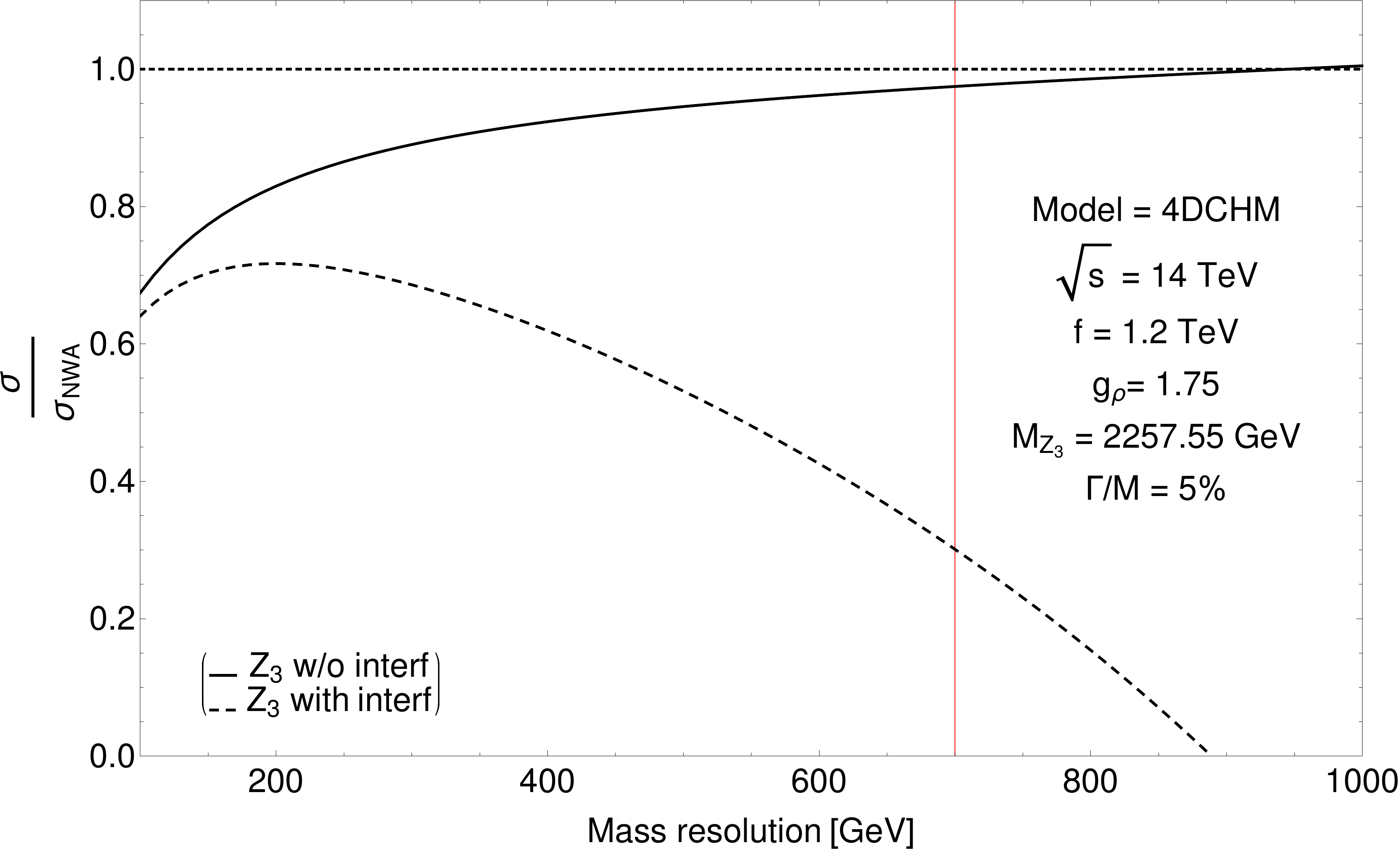}
\label{fig2b}
}
\caption{Ratio of the integrated cross section for the signal of a $Z_3$ boson in the 4DCHM over the NWA result. The dotted line is the $Z_3$ in NWA result, 
the solid line is the $Z_3$ contribution including FW effects and the dashed line is the $Z_3$ contribution including both FW and 
interference effects.
\subref{fig2a} We start the integration at 2 TeV and the result is plotted as a function of the resonance width over its mass (in percentage). 
\subref{fig2b} We plot the integrated cross section as a function of the symmetric integration interval around the peak. The vertical red line 
represents the CMS adopted optimal cut which keeps the interference and FW effects below the 10\% in the case of narrow single $Z^\prime$ 
models \cite{Accomando:2013sfa}.
}
\label{fig2}
\end{figure}

In fig. \ref{fig2}a we show the deviations from the NWA results, as a
function of the resonance width. In this particular case the
cross sections are obtained by integrating from 2 TeV to effectively
infinity. This means that we have not included a large part of the
negative interference contribution stemming from the low mass region.
Significant deviations from the NWA model are observed (dotted line
compared with the dashed line) and can in general be larger if cross
sections are integrated over all masses even leading to an overall signal
cross section that is negative.
This demonstrates that inclusion of FW and/or interference effects results in 
a non-trivial modification of the
integrated cross section that cannot be accounted for by a simple
rescaling of the NWA results, nor do are there cancellations between FW and
interference effects leading to anything close to the NWA results. This is in 
contrast to \cite{Pappadopulo:2014qza} where such effects are
claimed to negligible.

The 4DCHM is indeed dominated by interference effects. This makes this scenario a
representative example of a large variety of $Z^\prime$ models that share this particular feature 
like Extra Dimensional models \cite{Accomando:1999sj, Bella:2010sc} and the recently published Custodial Vector Model \cite{Becciolini:2014eba}.
Large negative contributions to the signal cross section
coming from the interference in the region just below the resonance
pole appear in many single $Z^\prime$ scenarios and can be particularly
large in models with multiple $Z^\prime$ resonances.
In these cases using the NWA is inappropriate, potentially 
leading to significant overestimations of the cross section that 
may be observed in experiments. 

The deviations from the naive NWA application with respect to the
consistent inclusion of FW and interference effects are summarized in
fig. \ref{fig2}b. Here we plot these differences as a function of the
integrating region around the $Z^\prime$ peak (that is integrating
$\pm$ the quoted mass resolution around the resonance pole).  

The red solid vertical line represents the integration region (suggested in
\cite{Accomando:2013sfa}) adopted by CMS in order to keep
the FW and interference effects below 10\% in the case of
narrow single $Z^\prime$ models. Clearly, for the 4DCHM this
prescription breaks down and we obtain substantial deviations from the
NWA predictions. Moreover, the picture becomes even worse as we
increase the resonance width. Again, scenarios where the (partially)
integrated signal cross section turns negative are not uncommon.

\section{Results for the complete 4DCHM}
Here we consider the complete 4DCHM where the $Z_2$ contribution
is also included, together with the $Z_3$. The other active resonance, the
$Z_5$, as already mentioned, is much heavier and thus difficult to produce,
ultimately giving a negligible contribution to the cross section
distribution in the invariant mass region we are interested in (around
the $Z_2$ and $Z_3$ poles). For these reasons and for ease of computation,
here, we neglect the $Z_5$ resonance.

\begin{figure}[t]
\centering
\subfigure[]{
\includegraphics[width=.45\textwidth]{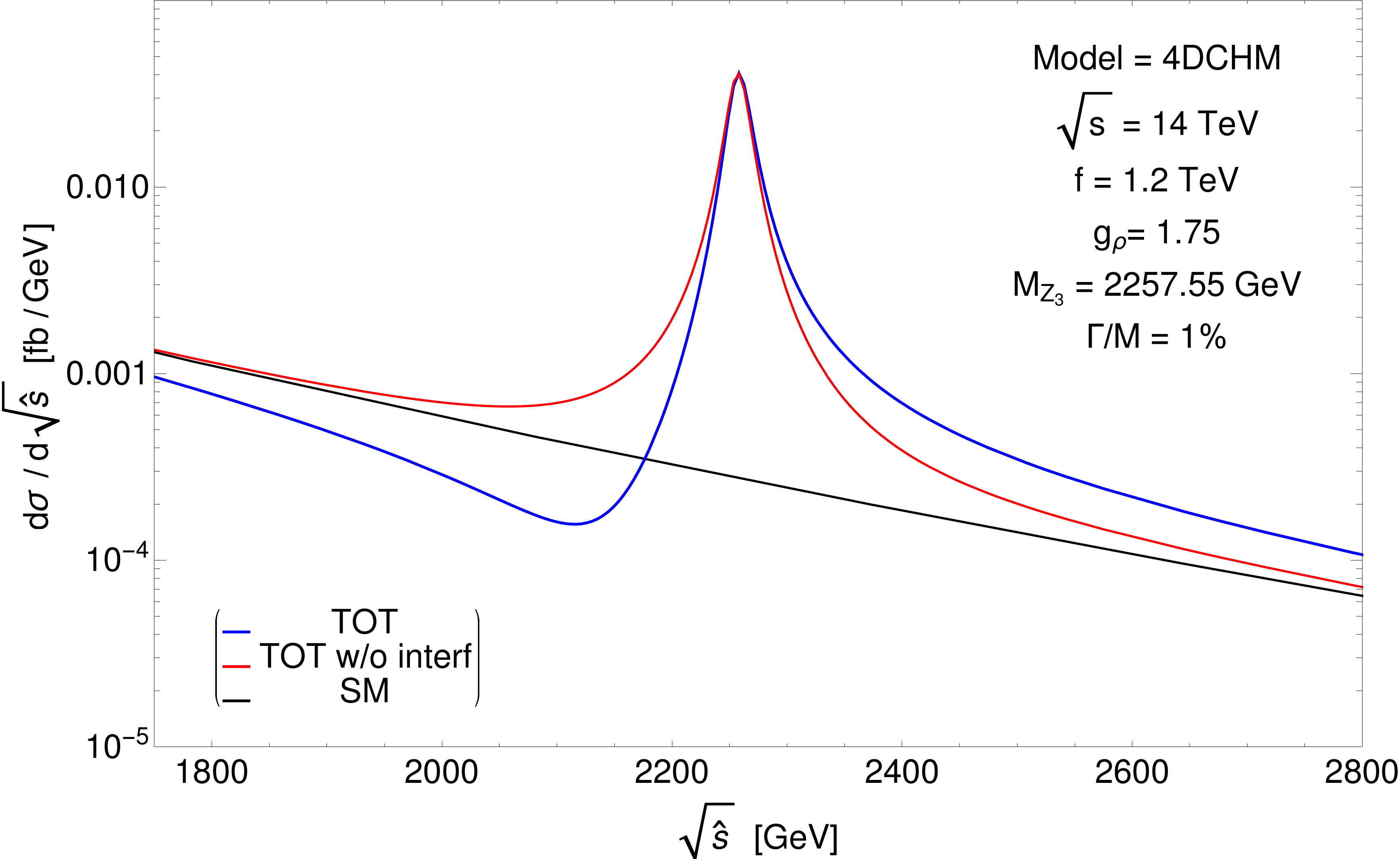}
\label{fig3a}
}
\subfigure[]{
\includegraphics[width=.45\textwidth]{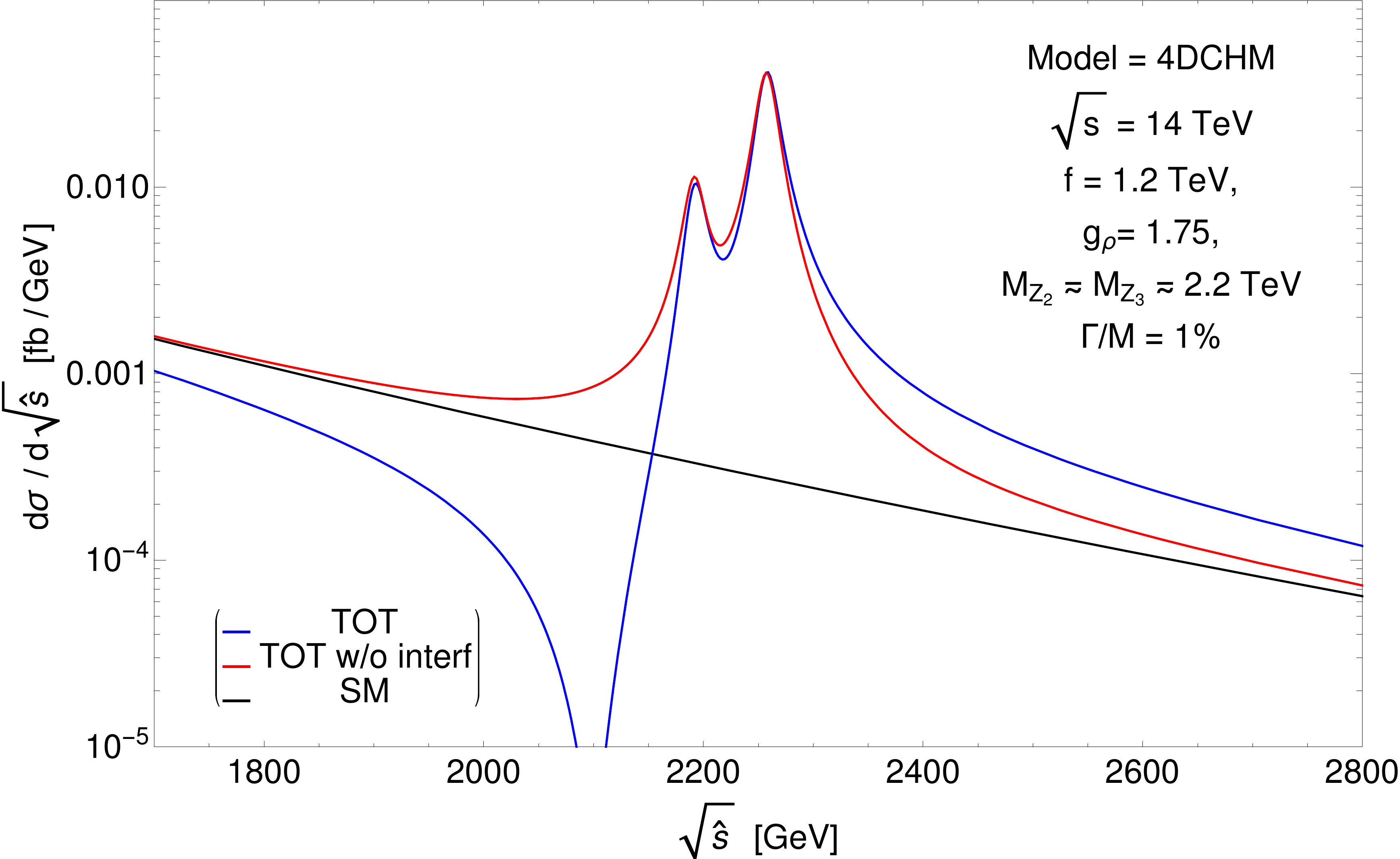}
\label{fig3b}
}
\caption{Cross section distribution for a specific point in the parameter space of the 4DCHM.
\subref{fig3a} We have considered only the $Z_3$ contribution and its interference with the SM background.
\subref{fig3b} The complete model is considered with the inclusion of the $Z_2$ boson exchange as well as its interference with the SM background and with the $Z_3$ boson.
}
\label{fig3}
\end{figure}

In fig. \ref{fig3} we compare the
cross section distribution in the case of the single $Z_3$ boson (a)
and of the complete 4DCHM (b).  As in the previous section
the negative contribution below the resonance peak coming from the
interference term spoils the NWA result: while already visible in the
case of the single $Z_3$ boson, it is even more evident in the
complete 4DCHM.

\begin{figure}
\centering
\includegraphics[width=.75\textwidth]{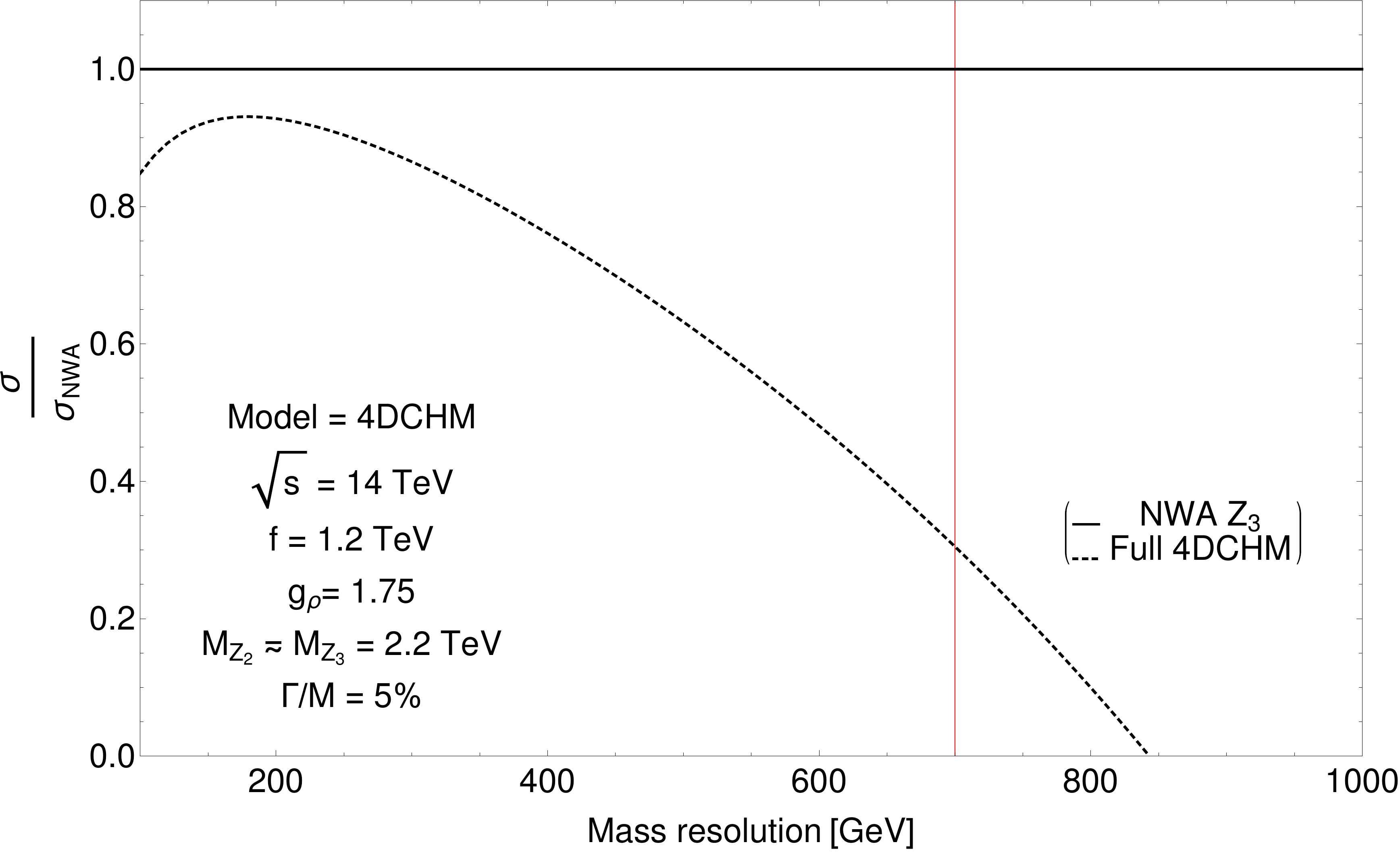}
\caption{Ratio of the integrated cross section in the full 4DCHM scenario over the NWA result as a function of the symmetric integration interval 
around the peak. The vertical red line represent the CMS adopted optimal cut which keeps the interference and FW effects below 10\% in the case 
of narrow single $Z^\prime$ models \cite{Accomando:2013sfa}.}
\label{fig4}
\end{figure}

Thus all the conclusions presented in the previous section are also valid
in the case of the complete model, where the deviations from
the naive NWA approach are even more remarkable. In order to show
these effects in the multi-$Z^\prime$ scenario, in fig. \ref{fig4} we
have repeated the exercise of the previous section, plotting the
integrated cross section as a function of the integration interval.

Last thing to mention is about the double peak structure that we
expect in the complete model case. 
In the particular example we examine here the double peak structure is
clearly visible at least before any detector resolutions are applied. 
Unfortunately this conclusion is not valid over the entirety of the 
model parameter space. The two resonances $Z_2$ and $Z_3$ can be very close
in mass and it is therefore often very difficult to separate them, especially as
their widths increase.

\section{Conclusions}
We have explored the phenomenology of the 4DCHM which is a realistic
and representative example of the class of multi-$Z^\prime$
scenarios. We have examined the consequences of large interference effects which
are not included in the NWA prescription and make this approach
invalid, even in the case where we only consider the dominant
resonant contribution, which comes from the $Z_3$ boson.
When we consider the complete model, that is, also including  the
contribution of the $Z_2$ boson as well as its interference (the
other active resonance, the $Z_5$, can be neglected as much heavier),
the picture gets even worse in terms of NWA validity. The total cross
section is significantly overestimated by the NWA approach. Any experimental 
interpretation of observations in the context of this type of model 
should take this dynamics into account and appropriately 
adapt the methodologies used. 

We have also seen that FW effects can be sizable, since there
are regions in the parameter space of the 4DCHM where the neutral
resonances can become very broad, due to the opening of new decay
channels into extra fermions. These features are quite common
in models with multiple-$Z^\prime$, therefore it may be difficult to
extract realistic bounds on a specific model using NWA assumptions.  These
drawbacks have been overlooked in several phenomenological analyses.

In summary, we have here reviewed the main effects which ought
to be taken into account in search analyses for multi-$Z^\prime$ scenarios. These are the FW and
interference effects, which generally manifest themselves as a
substantial (negative) dip below the usual $Z^\prime$ peak(s).
We expect this dynamics to be common to generic Composite Higgs Models.

Finite Width and Interference effects have already been taken into account by the CMS collaboration in a sophisticated and dedicated way for the 
$W^\prime$-boson search at the LHC \cite{Accomando:2011eu, Khachatryan:2014tva, CMS:2015vda}. The analysis of the LHC data at 8 TeV has shown that the extracted limits on the $W^\prime$ mass have 
changed sizeably. We suggest that a similar approach should be adopted for the multi-$Z^\prime$ search in Run II.

\section*{Acknowledgements}
\noindent
E. Accomando, J. Fiaschi, S. Moretti \& C. Shepherd-Themistocleous are supported in part through the NExT Institute. 

\bibliography{references}

\end{document}